\documentclass[twocolumn,prx,noshowpacs,floatfix,groupedaddress]{revtex4}
\usepackage{graphicx}
\usepackage{epstopdf}
\usepackage{amssymb, amsmath}
\usepackage{multirow}
\usepackage{dcolumn}
\usepackage{bm}
\usepackage{subfigure}
\usepackage{color}
\usepackage{ulem}
\usepackage{textcomp}

\newcommand{\COMMENTED}[1]{}

\begin{document}

\title{Infinite Variance in Fermion Quantum Monte Carlo Calculations}
\author{Hao Shi}
\author{Shiwei Zhang}

\affiliation{Department of Physics,
             The College of William and Mary,
             Williamsburg, Virginia 23187}

\begin{abstract}

For important classes of many-fermion problems, quantum Monte
Carlo (QMC) methods allow exact calculations of ground-state and finite-temperature properties,
without the sign problem. The list spans condensed matter, nuclear
physics, and high-energy physics, including
the half-filled repulsive Hubbard model, the spin-balanced atomic
Fermi gas, lattice QCD calculations at zero density with Wilson Fermions, 
and is growing rapidly as a number of problems 
have been discovered recently 
to be free of the sign problem. 
In these situations, QMC calculations are 
relied upon to provide definitive answers. Their results are instrumental to our ability 
to understand and compute properties in fundamental models important to multiple sub-areas in 
quantum physics.  
It is shown, however, that
the most commonly employed algorithms in such situations turn out to
have an infinite variance problem.
A diverging variance causes the estimated Monte Carlo statistical
error bar to be 
incorrect, which can render the
results of the calculation unreliable or meaningless.
We discuss how to identify the infinite variance problem.  An
approach is then proposed to solve the problem. The solution
does not require major modifications to standard 
algorithms, adding a ``bridge link" to the imaginary-time path-integral.
The general idea is applicable to a variety of situations where the infinite variance problem may be present.
Illustrative results are presented for the ground state of 
the Hubbard model at half-filling.

\end{abstract}


\maketitle

\section{Introduction}
\label{sec:intro}

Quantum Monte Carlo (QMC) methods refer to a large family of modern computational approaches to compute properties of interacting quantum-mechanical systems.
They are widely used in condensed-matter physics, nuclear physics, high-energy physics, and quantum chemistry. There are different flavors of QMC, all of which involve the 
use of
Monte Carlo (MC) sampling
techniques to evaluate 
some form of path-integrals representing the many-body ground-state wave function or finite-temperature partition function or action. Because of the size of the underlying Hilbert space in the quantum system, the dimension of the integrals 
involved
is often enormous, making it difficult or intractable for other computational approaches. 
QMC methods thus 
play a fundamental role in the study of quantum models and materials.

For a variety of boson systems \cite{PIMC:Ceperley:He4:RMP,PhysRevE.74.036701} and unfrustrated spin models \cite{PhysRevB.43.5950}, the integrand is a positive function which 
resembles the partition function 
in classical systems. 
The calculations then behave like classical MC simulations, albeit with added complexities and effectively in higher dimensions. 
A successful QMC calculations yields the expectation value(s) of the physical observable(s), with 
an estimate of the statistical error bar(s). The MC result is only meaningful when accompanied by a reliable error
bar, which provides a statistical measure of the range of the possible answer
with respect to the computed expectation value. 
 
For systems with fermions, the exchange symmetry
dictates that, in general, the integrand cannot be made all positive. A sign problem 
\cite{SIGNPPRB1990,PhysRevLett.94.170201,RevModPhys.73.33,PhysRevLett.83.2777}
then arises. 
This problem fundamentally changes the (low) algebraic scaling of the computational time 
with respect to 
system size or inverse temperature \cite{PhysRevLett.83.3116}, 
making the statistical noise in the computed results grow exponentially.
In order to remove the exponential scaling, approximations \cite{FNKalosJcp,DmcCeperleyJcp,RevModPhys.73.33,shiweiPRL1995} are generally needed which introduce a systematic
bias in the calculated results. 
Computations in fermion systems are thus often drastically harder than in boson systems,
and reliable results are much more difficult to achieve.

For 
important classes of fermion problems, however, the calculations can be formulated
to be free of the sign problem. 
Examples span multiple areas in physics, and range 
 from 
the half-filled repulsive Hubbard model for magnetism and possible spin liquid states \cite{Meng2010,sorella2012absence} to spin-balanced fermions with attractive 
interaction describing atomic Fermi gases  to 
Kane-Mele models \cite{PhysRevLett.106.100403} and spinless fermion models \cite{PhysRevB.91.241117,PhysRevB.91.235151}
for topological phases to 
zero-density lattice QCD calculations \cite{PhysRevD.10.2445,RevModPhys.82.1349,0954-3899-39-9-093002} with Wilson Fermions.
These calculations employ the 
determinantal QMC approach based on auxiliary-fields 
\cite{PhysRevB.31.4403,BSS1PhysRevD.24.2278,KooninSugiyama19861,Sorella1989,PhysRevB.40.506}.
By exploiting certain symmetries of the problems, 
the integrand in the
many-dimensional integral, despite fermion antisymmetry, can be made non-negative
in this method.
These classes of fermion problems are growing in number and in impact, as 
more problems are being discovered and more models are being proposed and studied
\cite{PhysRevB.71.155115,2015arXiv150605349W,PhysRevB.91.235151,PhysRevB.91.241117,HShi-SZhang-to-be-published} 
where the sign problem can be made absent in a similar manner.
In these situations, the QMC calculation is relied upon to provide definitive answers for our understanding of 
fundamental models or systems, much like in boson 
systems, unfrustrated quantum spin models,  or in classical MC simulations.  

In this paper, we show that 
the  commonly employed forms of the determinantal QMC approach,
as applied to such situations,
have MC variances that diverge. 
Since the MC statistical error is proportional to the variance,  
the divergence 
makes it impossible to 
obtain a correct estimate   
of the error bar, thereby rendering the MC results unreliable.
The results obtained by ignoring the problem can
 turn out to be reasonable, as we illustrate below.
However, 
the computation cannot internally determine whether this will be the case and, in a strict mathematical sense, 
the result is not meaningful without controlling the problem.
The extent of the problem can differ for different models, observables, and algorithms, 
but the fundamental problem appears to be 
generic in standard path-integral determinantal QMC calculations. 

We illustrate the infinite variance problem, discuss its origin, and examine ways to detect it. 
A method is then proposed to 
solve the problem, which 
is straightforward to implement within the standard algorithms.
The work here
provides a robust approach for 
all the situations mentioned above where standard determinantal QMC algorithms are
applied to  sign-problem-free fermion systems. Further, 
the ideas 
can be potentially useful in many other MC simulations (wherever  
the function being sampled might contain zero values).

The remainder of this paper is organized as follows. 
In Sec.~II, we illustrate the infinite variance problem in a normal determinantal QMC calculation in the Hubbard model. 
In Sec.~III, we summarize the formalism of determinantal QMC, focusing on ground-state calculations, to facilitate the 
ensuing analysis of the origin and the presentation of our solution for the variance problem. 
In Sec.~IV, we study the variance problem using a toy problem which can be thoroughly 
examined analytically. Then in Sec.~V, we present our solution to the infinite variance problem. 
This is followed by discussions of the general applicability of our approach and several technical aspects, 
in Sec.~VI. We end with summary and conclusions in Sec.~VII.

\section{Symptoms of the problem} \label{sec:symptoms}

\begin{figure}[htbp]
 \includegraphics[scale=0.5]{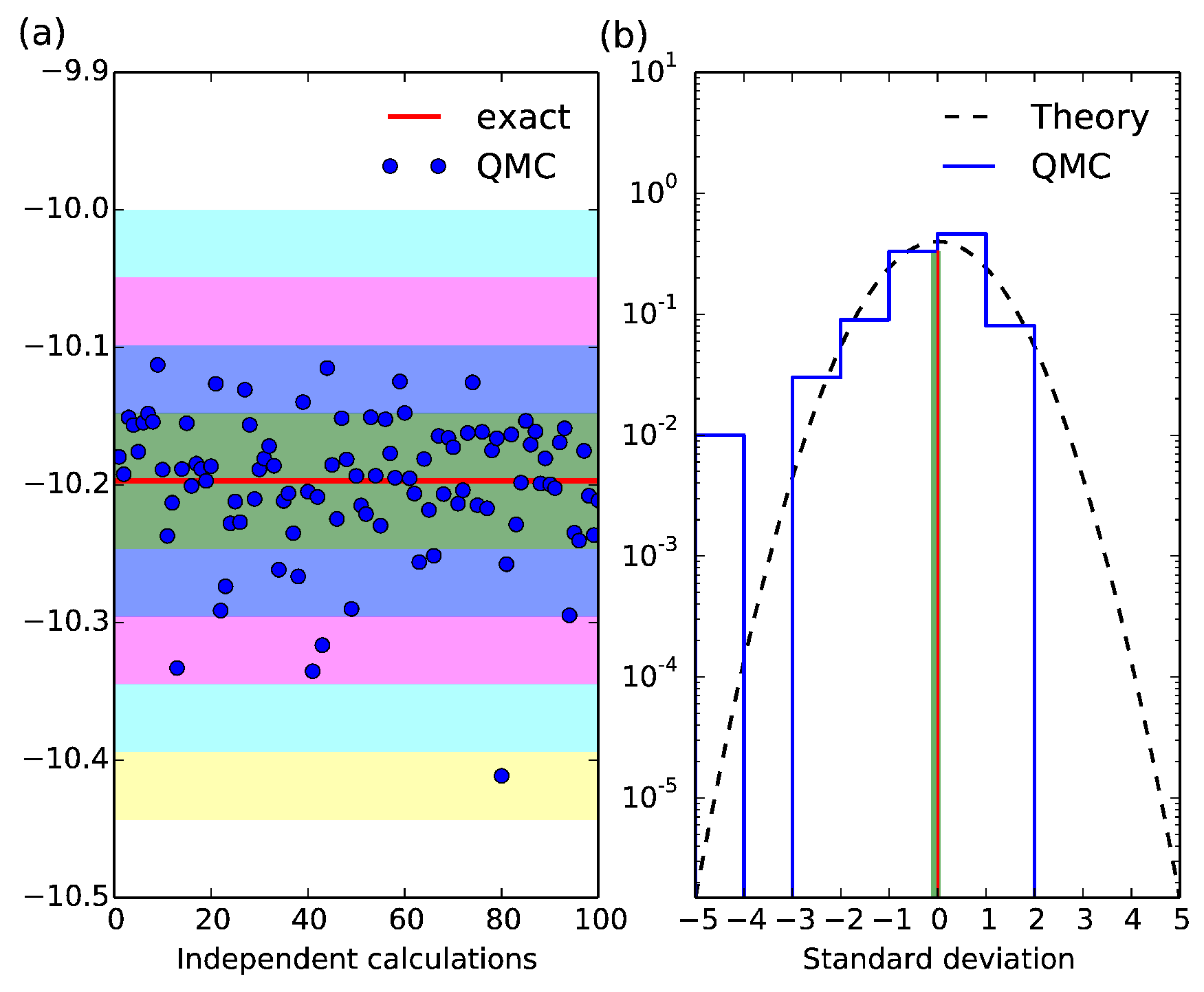}
 \caption{\label{fig:problem} 
          (Color online) Distribution of the computed 
          ground-state energies from
          $100$ independent determinantal QMC calculations. 
          In the left panel, each shade band indicates one standard deviation, as computed from the 100 data points, which 
          are plotted vs.~the (arbitrary) run index. In the right panel, the histogram of the 100 data points are shown 
          with a bin size of the computed standard deviation. For comparison, the theoretical Gaussian distribution 
          from the CLT is also shown. (A shift is applied on the horizontal axis so that the exact result is shown 
          at zero, the vertical red line.)
          The computed mean from the 100 data is shown by the vertical blue line, with its thickness indicating 
          the statistical error bar.
          Note the logarithmic scale on the vertical 
          axis. 
         } 
\end{figure}

In this section, we illustrate the infinite variance problem with calculations in the 
Hubbard model: 
\begin{equation}
\label{eq:Hub-H}
\hat{H}=-t \sum_{<i,j>,\sigma}^{N} c_{i\sigma}^{\dagger}c_{j\sigma}+U\sum_{i}^{N} n_{i\uparrow}n_{i\downarrow}.
\end{equation}
Here $N$ is the number of lattice sites, $\langle,\rangle$ denotes nearest neighbor, the operator $c_{i\sigma}^{\dagger}$ ($c_{i\sigma}$) creates (annihilates) a electron of spin $\sigma$ on the $i$-th lattice site, and $U$ is the interaction when
two electrons of opposite spins are on the same lattice site. We will assume that 
there are equal numbers of electrons of both spins: $N_\uparrow=N_\downarrow=N/2$. 

With a repulsive interaction ($U>0$), there is no sign problem on a bipartite lattice at half-filling. The reason for this,
as well as details of the standard determinantal QMC algorithm we employ, will be 
discussed in the next section. 
For  illustrative purposes, we have selected an arbitrary small system,  a $2\times 4$ supercell.
The characteristics of the results discussed and the underlying issues are general and independent of the
details of 
system or the calculation.
We compute the 
total ground-state energy of the system 
in 100 independent calculations. 
Each calculation 
carries out, by Metropolis MC sampling of the path-integral form, the  
imaginary-time projection from
an initial wave function taken to be 
the ground state of the non-interacting system.
The total imaginary time of the projection in each calculation is $\beta=81$, with $\Delta \tau =0.01$ (in units of $t$). 
After discarding an initial equilibration phase, we perform $50$ sweeps along the path 
measuring the energy between $0$ and $81t$ with an 
interval of $0.9t$ of imaginary time.

From 
standard analysis procedure, one obtains the expectation value from the
average of the 100 data points. 
The statistical error bar is given, based on the Central Limit Theorem (CLT),
by the standard deviation divided by $\sqrt{100-1}$.
Our final result is  $-10.199\pm 0.005$.
This implies that, for example, 
the probability that 
the exact result is
within one MC error bar of the computed expectation is $\sim 68.27\%$,  the probability
that it is outside of two
error bars is $4.55\%$, etc.
The exact result is $-10.197$ \cite{note-twistBC}, and
our results look quite reasonable.

However, as seen from Fig.~\ref{fig:problem}, 
the data exhibit several anomalies. 
The distribution of the MC data is not symmetric about the expectation value. 
One data point falls outside four standard deviations from the mean,
the probability for which should be less than $0.007\%$.  
Overall, the $\chi^2$ between the two distributions in the right panel of Fig.~\ref{fig:problem}, namely
the histogram
from binned data and the theoretical distribution
according 
to the CLT, 
is $342.1$, which indicates that it is highly unlikely that the two are consistent. 
The disagreement means that 
although our final result happens to be consistent with the exact result, 
the MC estimates of the mean and statistical error bar could have been  catastrophically wrong.

We have tested many different system sizes and interaction strengths, 
several different models, different forms of 
HS transformations, and measuring observables other than the energy,
to confirm the above  observations.
The behavior appears quite general in standard determinantal QMC calculations. 
In the next section we discuss the problem more formally 
and provide an explanation of its origin.
 
\section{Determinantal QMC formalism and the origin of the problem}
\label{sec:formalism}

We first  briefly outline the standard determinantal QMC method. 
For concreteness, we will focus on the ground-state algorithm. 
The finite-temperature, grand-canonical formalism, on which we will comment further in Sec.~\ref{sec:discussion},
shares many of the same features.
In most ground-state QMC approaches, 
the imaginary time operator $\exp(-\beta \hat{H}/2)$ is applied to an initial wave function $|\psi_T \rangle$. 
If $|\psi_T \rangle$ is not orthogonal with the ground state $|\psi_0 \rangle$, the process will converge to $|\psi_0 \rangle$ 
for sufficiently large $\beta$.
The expectation value of a ground-state observable or correlation function $\hat O$
can be measured by:
\begin{equation}
\label{eqn:projection}
 \langle \hat{O}\rangle =\frac{\langle \psi_T|\,\exp(-\beta \hat{H}/2)\,\hat{O}\,\exp(-\beta\hat{H}/2)\,|\psi_T \rangle}{ \langle \psi_T|\,\exp(-\beta \hat{H})\,|\psi_T \rangle}.
\end{equation}

In the standard ground-state determinantal QMC method, the projection operator is expressed 
as an integration of the one-body propagators 
\cite{AFQMC-lecture-notes-2013} via the use of 
the Trotter-Suzuki breakup and Hubbard-Stratonovich (HS) transformation
\begin{equation}
 \label{eq:Bx}
 e^{-\Delta \tau \hat{H}} =\int  p(x) \hat{B}(x) dx,
\end{equation}
where 
$\Delta \tau$ is a small time step, and 
$\beta=L \Delta \tau$.
The details of the functions on the right-hand side depend on 
the 
Hamiltonian and the particular choice of the HS transformation, 
but the  form in Eq.~(\ref{eq:Bx}),
holds generally.
The variable $x$, which is referred to as an auxiliary-field,  
is a many-dimensional vector and can be either continuous or discrete. (In the latter case the integral 
in Eq.~(\ref{eq:Bx}) is 
actually a sum.) The function $p(x)$ can be viewed as a probability density, and the one-body propagator 
has the general form: $ \hat{B}(x) =\exp(\sum_{i,j} b_{i,j}(x) c^\dagger_i c_j)$. Any one-body propagator of this form has 
the property 
\begin{equation}
\hat{B}(x) |\phi\rangle =|\phi'\rangle\,,
 \label{eq:Thouless}
\end{equation}
where $|\phi\rangle$ and $|\phi'\rangle$ are independent-particle fermion wave functions, i.e., Slater determinants. 
The orbitals in $|\phi'\rangle$  are related to those in $|\phi\rangle$ by a simple matrix product
using the matrix corresponding to the operator $\hat{B}(x)$.

The equation in (\ref{eqn:projection}) then becomes a multi-dimensional path integral in auxiliary-field space:
\begin{widetext}
\begin{equation}
 \langle \hat{O}\rangle= \frac{\idotsint dx_1 dx_2 \ldots dx_L\,p(x_1)\ldots p(x_L)\, \langle \psi_T|\,\hat{B}(x_1)\ldots \hat{B}(x_{L/2})\, \hat{O}\,\hat{B}(x_{L/2+1})\ldots\hat{B}(x_L)\,|\psi_T \rangle}
   { \idotsint dx_1 dx_2 \ldots dx_L\,p(x_1)\ldots p(x_L)\,\langle \psi_T|\,\hat{B}(x_1)\,\hat{B}(x_2)\ldots\hat{B}(x_L)\,|\psi_T \rangle}\,.
\label{eq:explicitPI}
\end{equation}
\end{widetext}
In Eq.~(\ref{eq:explicitPI}) we have inserted $\hat{O}$ in the middle of the path, as we had done in Eq.~(\ref{eqn:projection}). 
Of course a measurement 
can be made anywhere along the path provided it is sufficiently far away from either end to ensure that 
convergence to the ground state has been reached by the projection from $|\psi_T \rangle$. This does 
not impact the discussion on the variance problem, and we will use Eq.~(\ref{eq:explicitPI})
 when the explicit formula is needed, with no loss of generality. 

If the initial wave function $|\psi_T \rangle$ is chosen to be a Slater determinant, the propagation by each auxiliary-field path, i.e., each string of $\hat{B}$ operators, keeps it in the form of a single Slater determinant.
 For brevity let us introduce the following notation:
\begin{equation}
 |\phi^{\rm r} (\pmb {X}_{\rm r})\rangle \equiv \hat{B}(x_{L/2+1})\hat{B}(x_{L/2+2})\ldots\hat{B}(x_L)|\,\psi_T \rangle
 \label{eq:phi_r}
\end{equation}
and 
\begin{equation}
 \langle \phi^{\rm l} (\pmb {X}_{\rm l})| \equiv \langle \psi_T|\,\hat{B}(x_{1})\hat{B}(x_{2})\ldots\hat{B}(x_{L/2})\,
 \label{eq:phi_l}
\end{equation}
where the shorthand $\pmb {X}_{\rm r}$ and  $\pmb {X}_{\rm l}$ denote the collection of auxiliary-fields with indices from 
$L/2+1$ to $L$ (inclusive) and from $1$ to $L/2$, respectively.
Further, let us denote $\pmb {X}\equiv \{\pmb {X}_{\rm l},\pmb {X}_{\rm r}\}=\{x_1, x_2, \ldots , x_L\}$, 
and the product of probability densities as $P(\pmb {X})=\prod_{l=1}^L p(x_l)$.

The integrand in the denominator in Eq.~(\ref{eq:explicitPI}) is given by 
\begin{equation}
 f(\pmb{X})=P(\pmb {X})\, \langle  \phi^{\rm l} (\pmb {X}_{\rm l})| \phi^{\rm r} (\pmb {X}_{\rm r})\rangle\,,
\label{eq:f}
\end{equation}
where the inner product can be conveniently evaluated as the determinant of the product of the matrices corresponding to the "left" and "right" wave functions
\cite{AFQMC-lecture-notes-2013}.
Similarly, the integrand in the numerator is given by
\begin{equation}
 g(\pmb{X})=P(\pmb {X})\, \langle  \phi^{\rm l} (\pmb {X}_{\rm l})|  \hat{O} |\phi^{\rm r} (\pmb {X}_{\rm r})\rangle\,,
\label{eq:g}
\end{equation}
so that Eq.~(\ref{eq:explicitPI}) reduces to a generic form:
 \begin{equation}
  \label{eqn:integral}
  \langle \hat{O}\rangle =\frac{\int g(\pmb{X})\,d \pmb{X} }{ \int  f(\pmb{X})\,d \pmb{X} }.
 \end{equation}

For general fermion problems, the determinant in Eq.~(\ref{eq:f}) can be both positive and negative as 
a function of $\pmb {X}$ --- 
indeed it is complex for problems with realistic electronic interactions \cite{PhysRevLett.90.136401}. 
However, as mentioned earlier, in many important 
classes of problems, $f(\pmb{X})$ turns out to be non-negative. 
For instance, in the repulsive half-filled Hubbard Hamiltonian of Eq.~(\ref{eq:Hub-H}), there is no sign problem 
as long as we choose a $|\psi_T\rangle$ which ensures partial particle-hole symmetry.
Either the charge or spin form of the HS transformation can be used, with either discrete or continuous
auxiliary-fields \cite{PhysRevB.88.125132}. 
This is one example where $f(\pmb{X})$ can be written as  the square or complex conjugation product of two determinants.
More generally, in these sign-problem-free situations $f(\pmb{X})$  can often be thought of as
the determinant of a matrix whose eigenvalues appear in pairs, either degenerate real values of 
complex conjugates.

For any problem with $f(\pmb{X})\ge 0$, it is straightforward to sample the 
probability density function (PDF): $f(\pmb{X})/ \int  f(\pmb{X})\,d \pmb{X}$ 
by Metropolis \cite{assaad-lec-note} or branching random walks \cite{AFQMC-lecture-notes-2013} 
and use MC to evaluate
Eq.~(\ref{eqn:integral}):
\begin{equation}
\langle \hat{O} \rangle\doteq \biggl \langle \frac{g(\pmb{X})}{f(\pmb{X})} \biggr \rangle_{f}\,,
\label{eq:Oavg}
\end{equation}
where the average is with respect to 
the configurations sampled from $f(\pmb{X})$.
The estimator $g/f$ reduces to 
$\langle  \phi^{\rm l} (\pmb {X}_{\rm l})|  \hat{O} |\phi^{\rm r} (\pmb {X}_{\rm r})\rangle/\langle  \phi^{\rm l} (\pmb {X}_{\rm l})|  \phi^{\rm r} (\pmb {X}_{\rm r})\rangle$, 
which is conveniently evaluated by the corresponding 
Green functions if $\hat{O}$ is a one-body operator and by combinations of Green functions
via Wick's theorem if $\hat{O}$ is a two-body correlation function \cite{assaad-lec-note,AFQMC-lecture-notes-2013}.
This is the standard determinantal QMC approach.

 In order to measure the statistical error bar, one computes the variance:
 \begin{equation}
  \sigma_{\hat{O}}^2 =\frac{\int  \frac{g^2(\pmb{X})}{f(\pmb{X})} d \pmb{X}  }{ \int  f(\pmb{X})  d \pmb{X}} - \langle \hat{O}\rangle ^2
  \doteq \biggl \langle \bigg( \frac{g}{f} \bigg)^2\biggr \rangle_{f} - 
  \biggl \langle \frac{g}{f} \biggr \rangle_{f}^2\,,
  \label{eq:var}
 \end{equation}
 where on the right we have omitted the variable $ \pmb{X}$  but the averages have the same meaning 
 as in Eq.~(\ref{eq:Oavg}). (In practice the configurations sampled will have auto-correlations, 
 and one will need to re-block the measurements to obtain a reliable estimate, as is commonly done. 
 This is always done in our data analysis in the present work.
It does not affect the following discussions.)
 
 The variance in Eq.~(\ref{eq:var}), as given by the explicit formula on the left, can diverge  
 if $g(\pmb{X})$ remains non-zero when $f(\pmb{X})$ approaches zero. 
 More precisely, it diverges if a non-integrable singularity exists in $g^2/f$ 
 anywhere in the 
 space of the auxiliary-field paths.
 This can occur because $f$ is given by the overlap between two single Slater determinant wave functions,
 $|\phi^{\rm l}\rangle$ and $|\phi^{\rm r}\rangle$, 
 which are randomly evolving. 
 The existence of paths with  $f(\pmb{X})=0$ is related to the occurrence of the sign 
 problem in calculations of general fermion problems in this framework. 
 The symmetry which prevents the sign problem from occurring 
 in the sign-problem-free cases
 eliminates the part of the space where
 $f(\pmb{X})< 0$; however, this symmetry in general does not exclude $f(\pmb{X})=0$.
 In the example in Sec.~\ref{sec:symptoms}, both 
 $|\phi^{\rm l}\rangle$ and $|\phi^{\rm r}\rangle$ can be written as 
 $|\phi_\uparrow\rangle \otimes |\phi_\downarrow\rangle$, where $ |\phi_\downarrow\rangle$ can be made 
 equal (or complex conjugate) to 
 $|\phi_\uparrow\rangle$ under partial particle-hole transformation \cite{PhysRevB.31.4403}. This means that 
 $f(\pmb{X})$ can be written in the form of $|\langle \phi^{\rm l}_\downarrow|\phi^{\rm r}_\downarrow\rangle|^2$
  for any path $\pmb{X}$, and thus is non-negative.
  It does not mean, however, that 
$f(\pmb{X})$ cannot be zero, which occurs whenever $|\phi^{\rm l}\rangle$ and $|\phi^{\rm r}\rangle$ become orthogonal. 
This is inevitable, since they are  two independent single-Slater determinant wave functions controlled by 
separate stochastic paths $\pmb {X}_{\rm l}$ and $\pmb {X}_{\rm r}$, respectively.

 This divergence is the origin of the symptoms 
 seen in the calculation in Sec.~\ref{sec:symptoms}.  
 It causes the underlying variance of the calculation 
 to diverge. It is important to emphasize that
 the infinite variance problem is not caused by a path with $f(\pmb{X})=0$
 being encountered in an actual calculation.
 In the MC calculation points with $f(\pmb{X})=0$ are, of course, 
 never sampled.
 The expectation value $\langle g/f\rangle$ exists, and will converge to the correct answer.
 The infinite variance problem arises because paths are sampled 
 which come close to $f(\pmb{X})=0$. 
 Although the computed variance, using $\langle (g/f)^2\rangle$,  will always have a finite numerical value
 in each calculation, it will  show sporadic behaviors from simulation to simulation.
 The variance is an intrinsic property of the underlying PDF being sampled, 
 so the problem does not depend on which  sampling algorithm is used. 
 The more samples generated, the more likely the divergence will manifest itself. 
 Hence the computed error bar, which is obtained by $\sigma$ divided by the number of effective independent samples, 
 does not provide a reliable estimate of the MC statistical error.

\section{Illustration from a toy model}
\label{sec:example}

 \begin{figure*}[htbp]
 \includegraphics[scale=0.5]{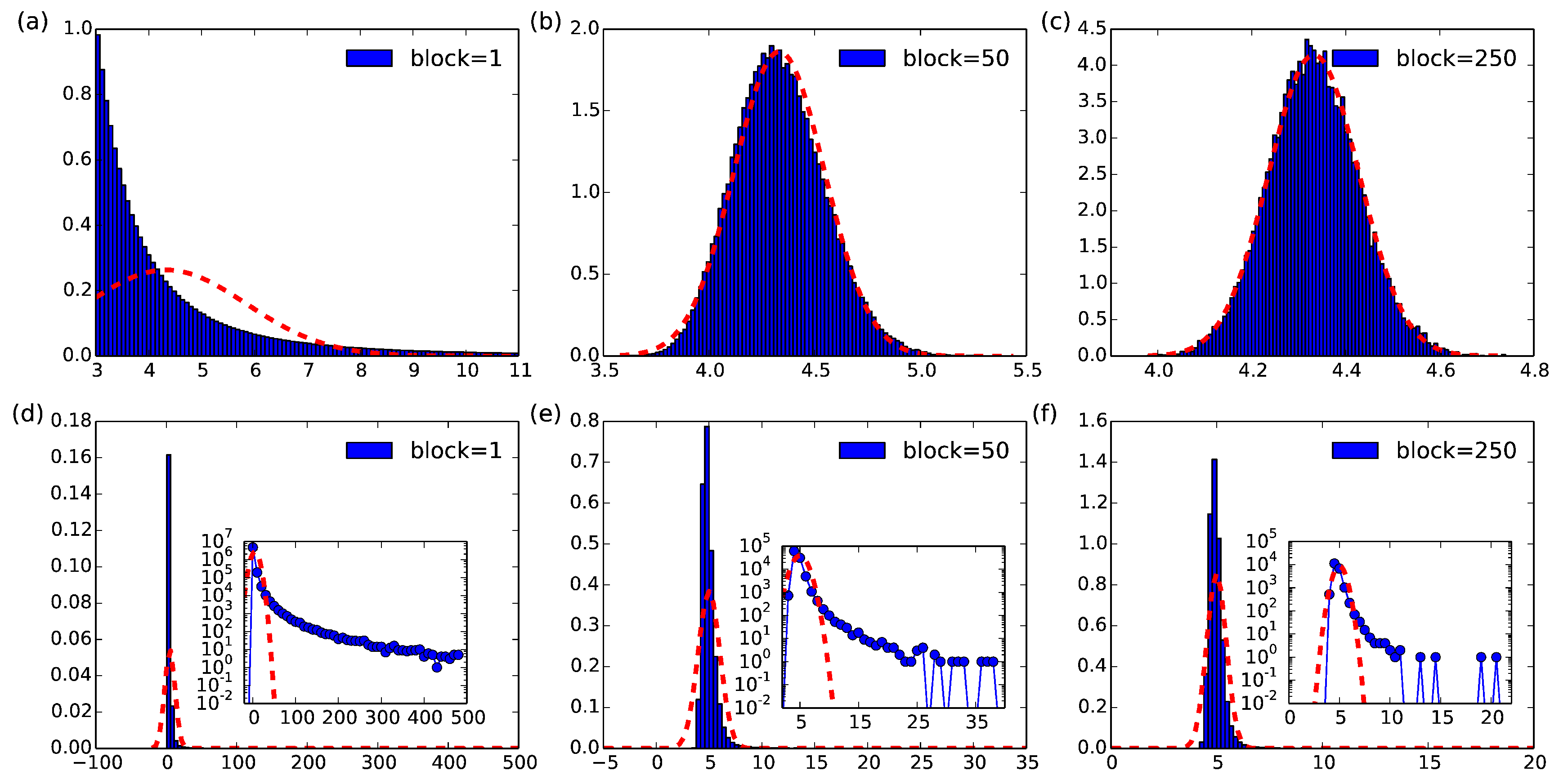}
 \caption{\label{fig:his-example} 
          (Color online) Normalized histograms of MC measurements for $\alpha=0.2$ (top row) and $\alpha=0.0$ (bottom row). 
          The calculations at each $\alpha$ are done with $5\times 10^6$ MC samples.
          Each histogram is obtained by grouping a different number ("block") of samples  
          together to make one entry of $\langle y(\alpha)\rangle$. 
          In the top row, they converge quickly to Gaussian distributions as "block" is increased,
          and reach agreement with  the red (dashed) curves, which indicate the Gaussian 
          as defined according to the CLT. 
          In the bottom row, in contrast, the histograms do not converge to Gaussians. A persistent discrepancy 
          is seen between the CLT prediction and the data. 
          The insets, which display the unnormalized histogram values (semi-log scale), highlight the long tail on the right. 
         } 
 \end{figure*}

 \begin{table}
 \caption{\label{tab:example} 
          MC results for the toy problem in Eq.~\ref{eqn:example} at $\alpha=0$. 
          The PDF $2x$ is sampled on $(0,1)$. 
          The MC statistical error bars (one standard deviation) are estimated from 100 independent runs.
         }
 \begin{ruledtabular}
 \begin{tabular}{cccc}
 Sample size & Computed value $\langle y(0)\rangle$ & Error bar\\
 \hline
 5000 &5.0064&0.0089 \\
 20000&4.9939&0.0047 \\
 80000&4.9997&0.0026\\
 320000&5.0011&0.0014\\
 1280000&5.0021&0.0009
 \end{tabular}
 \end{ruledtabular}
 \end{table}

  \begin{figure*}[htbp]
 \includegraphics[scale=0.42]{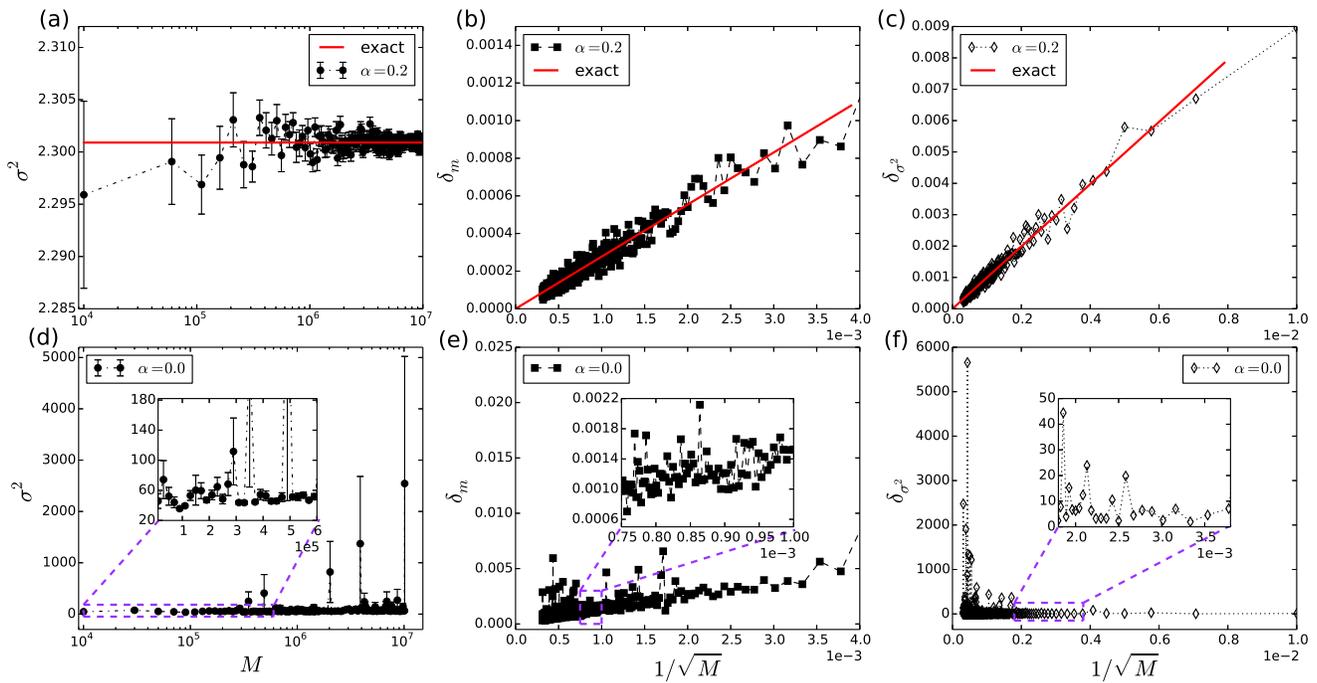}
 \caption{\label{fig:std-example} 
          (Color online) Comparison of finite and infinite variance calculations.
          The variance, statistical error on the expectation value, and the statistical error on the computed variance
          are shown for $\alpha=0.2$ (top row) and $\alpha=0$ (bottom row). The statistical errors are estimated from 
          30 independent MC runs. For $\alpha=0.2$, the computed variance remains consistent with analytic results as 
          the sample size $M$ is varied, and the computed statistical errors on the variance and on the expectation values 
          decreases as $1/\sqrt{M}$. For $\alpha=0$ the MC variance shows 
          increasing fluctuations as $M$ is increased,
          and does not converge. The statistical errors do not decrease following $1/\sqrt{M}$, as is especially evident in 
          the error on the variance.
         } 
 \end{figure*}
 
 In this section
 we illustrate several key aspects of the infinite variance problem by
 studying a toy problem.
 Let us consider the following expression 
 involving 
 simple one-dimensional integrals,
 \begin{equation}
 \label{eqn:example}
  y(\alpha) =\frac{\int_\alpha^1  (x+2 )\,d x}{ \int_\alpha^1 x\,d x}\,,
 \end{equation}
 where $\alpha\in[0,1)$ is a parameter which we will vary. 
 Eq.~(\ref{eqn:example}) can be viewed as a special case of Eq.~(\ref{eqn:integral}), with $f(x)=x$ and $g(x)=x+2$. 

 Mimicking the QMC calculations, we will 
 choose to sample the PDF  
 $x/( \int_\alpha^1 x\,d x) $ and evaluate $y(\alpha)$ by MC, i.e., the expectation 
 of $(x+2)/x$ from the samples.
 The exact value is $y(\alpha)=(5+\alpha)/(1+\alpha)$. 
 The variance is 
 \begin{equation}
 \begin{split} 
 \sigma_{y(\alpha)}^2 &=\frac{\int_\alpha^1 (x+2)^2/x\,d x } { \int_\alpha^1 x\,d x  }-y^2(\alpha) \\
                      &=-\frac{8\ln \alpha}{1-\alpha^2}-\frac{16}{(1+\alpha)^2}\,.
 \end{split}
 \label{eq:toy-var}
 \end{equation}
 As $\alpha\xrightarrow{}0$, the expectation 
 $y(\alpha)\xrightarrow{}5$ is well defined, while the variance diverges as $\sigma_{y(\alpha)}^2  \xrightarrow{}-8 \ln(\alpha)$.
 
 This divergence is not straightforward to detect 
 in the MC calculation.
 The logarithmic divergence is a consequence of samples landing closer to $f(x)=0$ 
 (i.e., $x$ being near the origin).
 Statistically
 this occurs  more as the sample size grows. On the other hand, the 
 standard deviation of the computed expectation value, in the absence of the divergence, will decrease
 as the square root of the sample size. In the competition between the two trends, 
 the logarithm is slower so the latter dominates. Table~\ref{tab:example} displays the result obtained from actual MC 
 calculations at $\alpha=0$.
 The expectation values are obtained from  averaging 100 independent runs,  each with the specified sample size,
 and the error bar is estimated by the standard deviation of the 100 results divided by $/\sqrt{99}$.
 Similar to the situation in the Hubbard model in 
 Sec.~\ref{sec:symptoms}, the results look reasonable at first glance. 
 The error bar is seen to decrease as the sample size is increased, roughly as the square root,
 although the largest calculation gives a result which is away from the correct answer by more than 
 two error bars. 
 
 In Fig.~\ref{fig:his-example} we examine the behavior of the calculations more closely, and compare it to 
 that of a case with no variance problem ($\alpha\neq0$).
 In each calculation a total of $M$ samples are drawn from the PDF. We group the samples into blocks
 each with $M_b$ samples, and compute the MC estimate of $y(\alpha)$  
 for each block. These are 
 entries for the histogram with "block" number $M_b$. Thus the first histogram in the top row 
 contains $M$ entries of  
 $(x+2)/x$, each computed at an $x$ value sampled from the PDF $f(x)\propto x$, with
  $x\in (\alpha,1)$.
 In the next histogram, each entry is obtained from an average 
 $\langle (x+2)/x \rangle $ of $M_b=50$ samples, and there are $M/M_b$ entries. 
 This procedure of re-blocking or re-binning is common in QMC calculations where auto-correlation
 is present. (If successive MC samples are not correlated, different ways of re-blocking will lead to 
 statistically equal 
 error estimates.) For each re-blocking step, the variance between the 
 block means can be computed numerically, following the right-hand side of Eq.~(\ref{eq:var}) 
 (with the block mean values replacing $g/f$) and averaging over all the blocks. 

  \begin{figure*}[htbp]
 \includegraphics[scale=0.50]{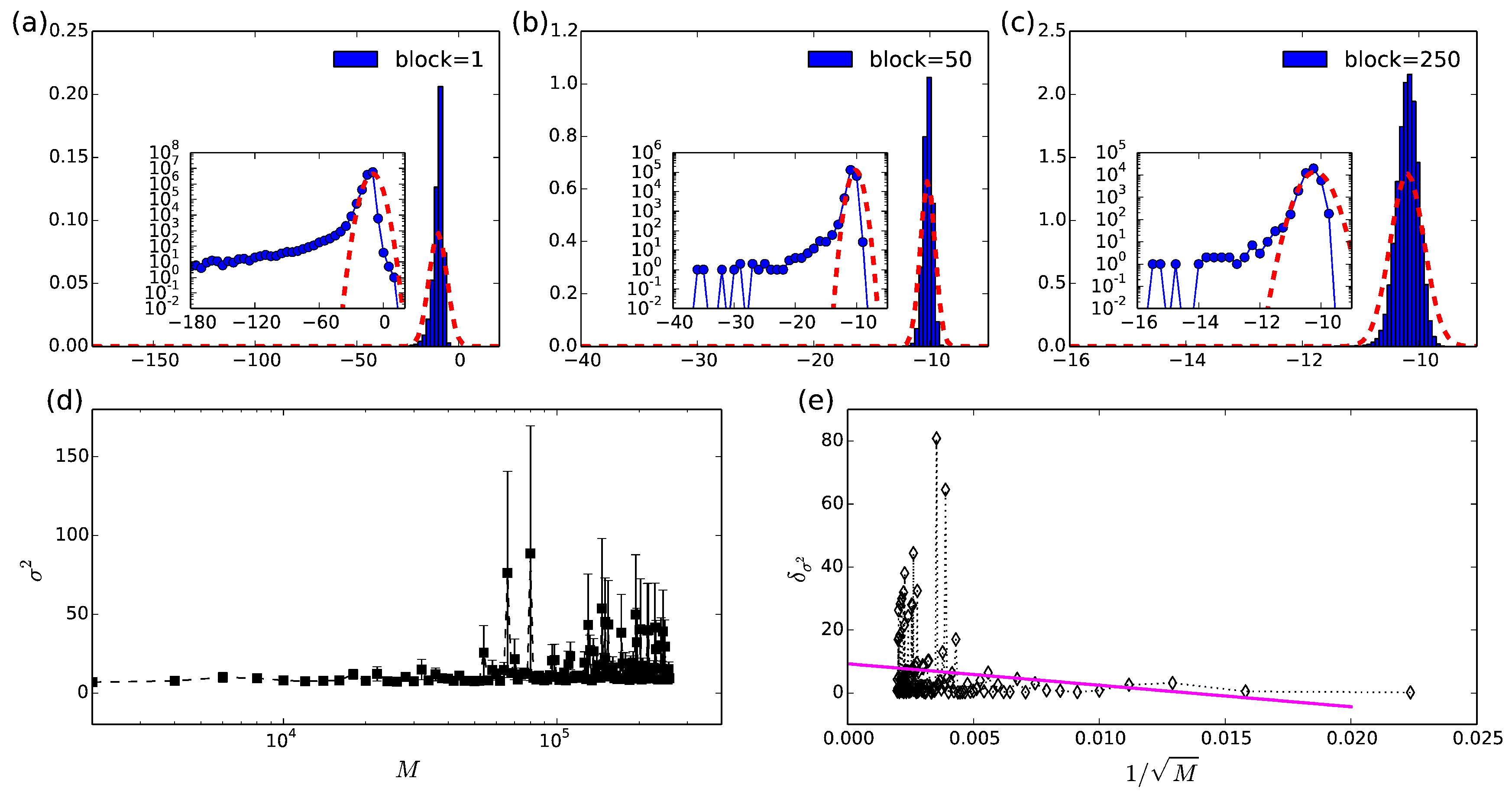}
 \caption{\label{fig:block-afqmc-infinite}
  (Color online) Detection and further analysis of the infinite variance problem in the Hubbard model calculation in
  Sec.~\ref{sec:symptoms}. The top panel shows a re-blocking analysis similar to that in Fig.~\ref{fig:his-example}. 
  The histograms of the computed ground-state energy do not converge to Gaussians and do not follow the CLT.
  In the bottom panel, the computed variance and the statistical error on the variance are shown vs.~sample size, 
  similar to Fig.~\ref{fig:std-example}. The variance does not converge to a finite value. Its error bar 
  grows with sample size in contrast with the expected $1/\sqrt{M}$ decay. (The magenta line is a linear fit.)
 } 
\end{figure*}

 As seen from the top row in Fig.~\ref{fig:his-example}, for $\alpha=0.2$ the behavior is 
 consistent with what is expected from the CLT. As $M_b$ is increased, the histograms converge to Gaussian 
 distributions given by  
 the overall mean and the standard deviations computed from the entries.
 For $\alpha=0$, however, the behavior is different. The histograms do not converge to Gaussian distributions with re-blocking. 
 A persistent 
 tail is present, 
 and the standard deviations and the MC error estimates obtained 
 according to the CLT do not give 
 a correct description of the actual data.

 Figure~\ref{fig:std-example} further analyzes the behaviors of the variance. 
 For each $\alpha$, we compute the variance and the expectation value systematically 
 for increasing sample sizes. In other words, a sequence of 
 $\langle y(\alpha)\rangle$ and $\sigma_{y(\alpha)}^2$ 
 are obtained as we vary the number of samples, $M$, used in Eqs.~(\ref{eq:Oavg}) and (\ref{eq:var}).
 To estimate the statistical errors on 
 $\langle y(\alpha)\rangle$ and $\sigma_{y(\alpha)}^2$  
 for each choice of $M$, 
 we carry out 30 independent MC calculations and compute the corresponding standard deviations.
(Note that this applies to any observables in any QMC calculations. An estimate of the error bar can always be 
obtained by repeating the calculations with different random number seeds a number of times 
and computing the standard deviation of the corresponding observable from them.)
 We see from the first panel in the figure that, at $\alpha=0.2$, the computed variance agrees with the exact value of 
 $\sigma^2=2.30087$ from Eq.~(\ref{eq:toy-var}), regardless of the sample size. The error bar on the computed 
 variance decreases with sample size. Indeed the error bar is proportional to $1/\sqrt{M}$ as shown in the last panel in the top row.
 Similarly, the statistical errors on the computed expectation value agree with $\sigma/\sqrt{M}$, 
 as shown in the second panel. For $\alpha=0$, the situation is different. 
 Though a well defined expectation  value still exists,
 the computed variance does not show convergence with increasing sample size. Large fluctuations
 are seen at large $M$ on the computed statistical errors of both the expectation value and, especially, 
 the variance. This is understandable, since larger $M$ makes it 
  it more likely to have samples which land ever closer to the origin.

 The toy problem is of course rather artificial.
 To what extent it captures the 
 characteristics of determinantal QMC is not immediately clear. Because of the non-orthogonal
 and over-complete nature of  $|\phi^{\rm l}\rangle$ and  $|\phi^{\rm r}\rangle$, less is known about the behavior of $f(x)$ and 
 how it approaches zero than that of wave functions written in coordinate space (which tend to vanish linearly at the node) \cite{Umrigar-variance}.
 In Fig.~\ref{fig:block-afqmc-infinite} a similar analysis is performed on the Hubbard model calculations described in 
 Sec.~\ref{sec:symptoms}. In the top panels histograms of the computed ground-state energy are 
 shown from re-blocking, again with the inset showing the long tails (which are on the left  since the energy is negative here). 
 In the bottom panels, the variance is computed with increasing MC sample size, following a similar procedure 
 to that used in Fig.~\ref{fig:std-example}. The estimated statistical error on the computed variance shows
 large fluctuations and does not resemble a $1/\sqrt{M}$ behavior. 
 As we see there is a 
 striking similarity between the behaviors of the real determinatal QMC calculations and the toy model.

\section{Solution}
 \label{sec:solution}

 Conceptually it is straightforward to avoid the infinite variance problem.
 One should modify the PDF which is sampled by MC so that it is non-zero in the entire configuration space (or 
 at least find one that only leads to an integrable singularity in the estimator).
 One example would be to shift the PDF, i.e., to sample 
 \begin{equation}
 f'(\pmb{X})=\frac{ f(\pmb{X})+ \eta }{\int [\,f(\pmb{X})+ \eta\,]\,d\pmb{X} }\,,
 \label{eq:fnew-shift}
 \end{equation}
 where $\eta$ is a small constant. One could also modify $f$ to set a minimum value such as
 $f'(\pmb{X})\propto \max\{f(\pmb{X}),\eta\}$.
 Yet another example would be to sample
 \begin{equation}
 f'(\pmb{X})\propto f(\pmb{X})+  \gamma |g(\pmb{X})| \,,
 \label{eq:fnew-mix}
 \end{equation}
 where $\gamma$ is a constant which can be tuned to minimize the variance of the desired 
 expectation value or a set of expectation values. 
One example is the worm algorithm \cite{PhysRevE.74.036701}, which uses this form as an elegant way
 to expand the sampled
 phase space beyond that defined by $f$.
 Under the new PDF $f'$,  the observable in Eq.~(\ref{eqn:integral}) can be estimated by 
 computing the integrals 
 in the numerator and denominator separately,
 \begin{equation}
 \langle \hat{O} \rangle=\frac{  \langle g(\pmb{X}) / f'(\pmb{X}) \rangle_{f'} }
                              {  \langle f(\pmb{X}) / f'(\pmb{X}) \rangle_{f'} }\,,
 \end{equation}
 where the averages are with respect to samples from the new PDF $f'(\pmb{X})$ as indicated. 
 These and related tricks have been used in different contexts
 \cite{ceperley:releasednode,releasesorella,Mirror-Potential-PhysRevC.32.1735}
 where a zero needs to be avoided in the function being sampled. 
 
 Any of the choices above would solve the toy problem 
 of Eq.~(\ref{eqn:example}). In realistic sign-problem-free QMC calculations, however,
 these approaches in general do not work well. 
 The function $f(\pmb{X})$ in these cases
 tends to span an enormous range. For example, we observe that the unnormalized $f(\pmb{X})$ can vary from $\exp(-50)$ to $\exp(50)$
 during a typical simulation in a lattice of moderate size.
 The range grows exponentially with system size (physical size or imaginary time/inverse temperature).
 This makes it difficult to choose a ``suitable'' value of $\eta$, which can 
 depend sensitively on the specific calculation. The choice can be either too small (no effect on reducing 
 the variance) or too large (inefficient sampling in a large part of the configure space and hence large variance).
 A reasonable choice for one
 can become ineffective for a different calculation (different physical, or even run, parameters). 
 In principle the approach in Eq.~(\ref{eq:fnew-mix}) could  work better if a suitable $g(\pmb{X})$ is found. 
 For example, we tested the case $\hat{O}=\hat{H}$ in the function in Eq.~(\ref{eq:g}).
 This was difficult to implement and it slowed down the computation significantly. 
 If one keeps  the measurement of $\hat{H}$ at a fixed location 
 on the path, say, at $l=L/2$, one has to re-compute large segments of the path for a two-body 
 expectation for every update, which is done in sweeps across the path. If one allows the position $l$ 
 to vary, the effective function in  the PDF is $g(\pmb{X}, l)$, for which detailed balance is less straightforward to maintain.
 
 Here we propose a simple solution to overcome the infinite variance problem which requires minimal modifications 
 to the standard algorithm. From Eq.~(\ref{eq:f}), let us introduce an intermediate function:
 \begin{equation}
 \digamma(\pmb{X})=
 P(\pmb {X})\, \langle  \phi^{\rm l} (\pmb {X}_{\rm l})|  e^{-\Delta \tau \hat{H}} |\phi^{\rm r} (\pmb {X}_{\rm r})\rangle\,.
 \label{eq:soln-F}
 \end{equation}
 We then define a new PDF to be used in the MC:
 \begin{equation}
 f'(\pmb{X};x')\propto
 P(\pmb {X})\, \langle  \phi^{\rm l} (\pmb {X}_{\rm l})|\,p(x')  \hat{B}(x')\,|\phi^{\rm r} (\pmb {X}_{\rm r})\rangle\,,
 \label{eq:soln-fp}
 \end{equation}
 which contains an extra auxiliary-field $x'$. 
 The function $\digamma(\pmb{X})$ implicitly depends on the location $l$ where the propagator $e^{-\Delta \tau \hat{H}}$ is inserted.
 The new PDF, on the other hand, does not distinguish where $x'$ is inserted. It is simply the PDF that 
 lives in a larger auxiliary-field space, identical to a path integral with $(L+1)$ time slices.
 Using Eq.~(\ref{eq:Bx}), we obtain that
 \begin{equation}
  \digamma (\pmb{X})=C \int f'(\pmb{X};x')\,dx'\,,
 \label{eq:soln-relation-F2f}
 \end{equation}
 where $C$ is a normalization constant (which will not need to be determined in the calculation).
  
 We can now write the original expectation value  in Eq.~(\ref{eqn:integral}) as
 \begin{equation}
 \langle \hat{O} \rangle=\frac{  \iint  \frac{g(\pmb{X})}{\digamma(\pmb{X})}\,f'(\pmb{X};x')\,dx'\,d\pmb{X} } 
 {  \iint  \frac{f(\pmb{X})}{\digamma(\pmb{X})}\,f'(\pmb{X};x')\,dx'\,d\pmb{X} }\,.
 \label{eq:soln-O-bridge}
 \end{equation}
 The identity is easily verified by performing the integrals over $x'$, using Eq.~(\ref{eq:soln-relation-F2f}). 
 This leads to the MC estimator
 \begin{equation}
 \langle \hat{O} \rangle=\frac{  \langle g(\pmb{X}) / \digamma(\pmb{X}) \rangle_{f'} }
                              {  \langle f(\pmb{X}) / \digamma(\pmb{X}) \rangle_{f'} }\,,
 \label{eq:soln-O-avg}                             
 \end{equation}
 where the average is with respect to the 
 PDF $f'(\pmb{X};x')$, which is sampled in the expanded space of auxiliary-field paths
 containing an additional time slice. The basic idea of the 
 new algorithm is thus:  
 \begin{enumerate}
 \item Set up the calculation with one more time slice than originally needed.  
 \item Update the entire path of $(L+1)$ time slices as usual.
 \item Whenever a measurement is made, the time
  slice where the measurement takes place is the ``extra'' time slice, which we shall refer to as the ``bridge''
  link. Its auxiliary-field configuration $x'$
  should be ignored, i.e., the corresponding $B(x')$ should be excluded in forming $f(\pmb{X})$,
  $g(\pmb{X})$ and $\digamma(\pmb{X})$. 
 \end{enumerate} 
 
 The ``bridge'' link is thus dynamic, moving along the path with the update sweeps. 
 This is a crucial difference from the approach
 of Eq.~(\ref{eq:fnew-mix}).
 Note that the integrals in Eq.~(\ref{eq:soln-O-bridge}) are automatically evaluated by MC when we
 perform the sampling in the expanded space of $\{\pmb{X},x'\}$ and ignore $x'$ in step (3). 
 Computing  $\digamma(\pmb{X})$ in Eq.~(\ref{eq:soln-O-avg}) requires 
 the expectation value of $\exp(-\Delta \tau \hat{H})$. We do so by expanding it 
 in terms of $\Delta \tau$. In most calculations this was done up to second order, which we 
 found 
 to be 
 sufficiently accurate. We discuss this point further in Sec.~\ref{sec:discussion}.
 
 The purpose of the intermediate function $\digamma$ is to remove any singularities in the expectations
 in Eq.~(\ref{eq:soln-O-avg}), without having to introduce a PDF that would decrease sampling efficiency drastically 
 or increase the complexity of the algorithm substantially. 
 The form of the PDF should scale properly to the thermodynamic limit, and its performance should 
 remain consistent  
 as system size and imaginary time length are varied.
 These are accomplished with 
 the form in Eq.~(\ref{eq:soln-F}), for a broad class of problems. 
 It is easy to see that the function $\digamma(\pmb{X})$ removes the zeros present in $f(\pmb{X})$. 
  From Eqs.~(\ref{eq:soln-relation-F2f})
 and (\ref{eq:soln-fp}), $\digamma(\pmb{X})$ 
 is a linear combination (over an infinite/large number of auxiliary-fields $x'$) of terms of 
 the form $\langle  \phi^{\rm l} (\pmb {X}_{\rm l})|\, \hat{B}(x')|\phi^{\rm r} (\pmb {X}_{\rm r})\rangle$.
 Each term in the integral/sum is non-negative.
 If the overlap between $\langle \phi^{\rm l} |$ and a single determinant 
 in the sum, $\hat{B}(x')|\phi^{\rm r}\rangle$,
 is zero for a particular $x'$, there will be different 
 random values of $x'$ which will give non-zero 
 contributions in the sum.
 
 For the energy, the estimator $g/\digamma$ in the numerator in Eq.~(\ref{eq:soln-O-avg})
 has the form $\langle  \phi^{\rm l}|{\hat H}|\phi^{\rm r}\rangle/\langle  \phi^{\rm l}|e^{-\Delta\tau {\hat H}}|\phi^{\rm r}\rangle$. It is easy to see that, to leading order in $\Delta\tau$,
  this is bounded by $-1/\Delta\tau$ (relative to the mean or trial energy). 
 It is worth emphasizing that the situation here is fundamentally different from that in 
 diffusion Monte Carlo \cite{RevModPhys.73.33} or in phase-free AFQMC \cite{PhysRevLett.90.136401}
 where one could encounter occasional walkers with large local energies.
 In those cases there is no infinite variance problem, as we further discuss in Sec.~\ref{sec:discussion}.
 To control the spurious fluctuations, one may apply a cutoff of ${\mathcal  O}(1/\sqrt{\Delta\tau})$
 on the local energies
  \cite{Umrigar1993,Wirawan-betatin-PhysRevB.80.214116} or use an estimate of the integral of the local energy over the time step \cite{Umrigar1993}. The key distinction is that there the problem 
  has a well-defined limit as 
  $\Delta\tau \rightarrow 0$,
  while here any artificial bounds applied on the local energy will 
  give back the infinite variance problem as one attempts to relax or extrapolate with the bound to remove the bias it introduces.

\begin{figure}[htbp]
 \includegraphics[scale=0.35]{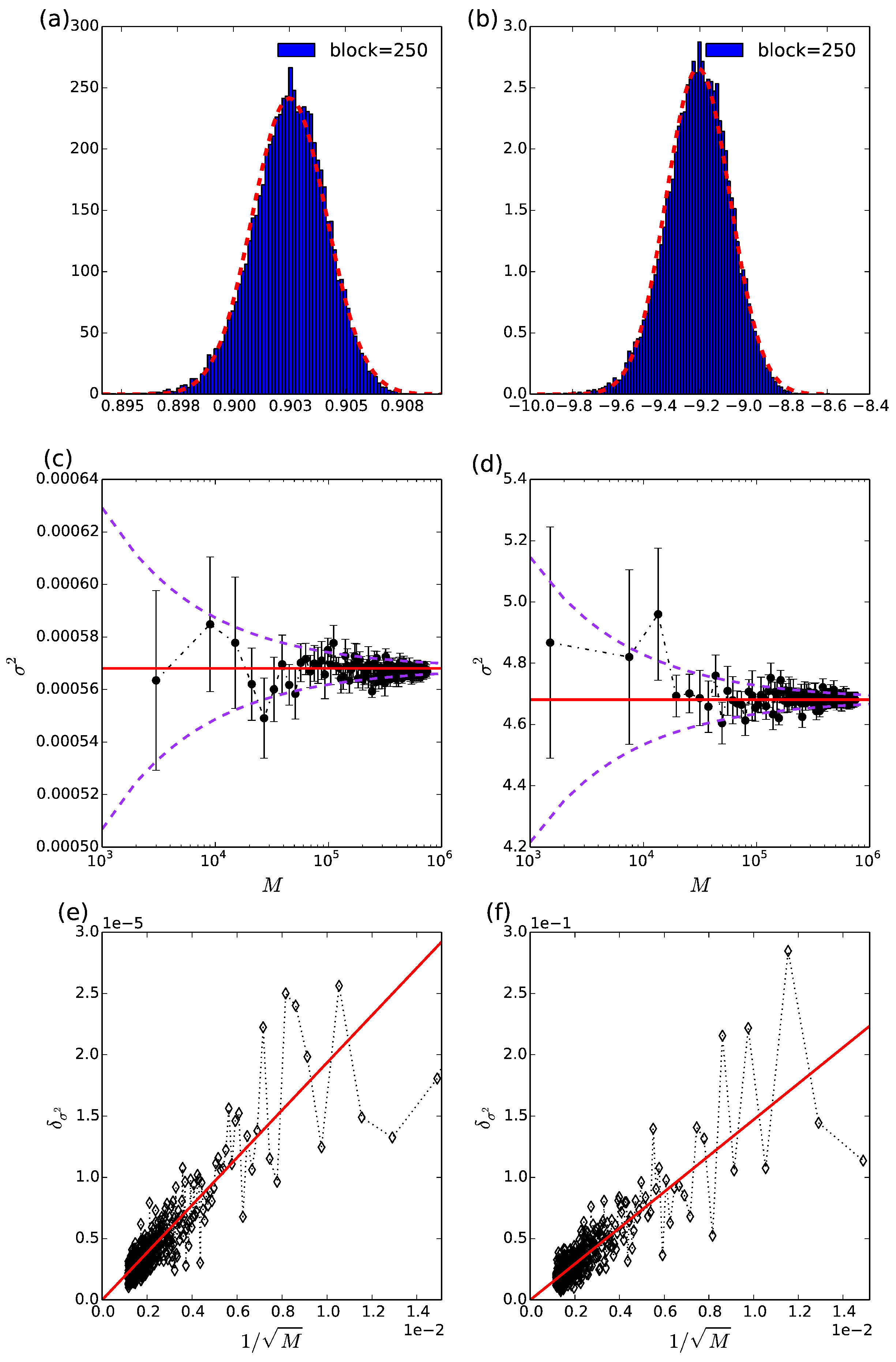}
 \caption{\label{fig:block-afqmc-finite} 
          (Color online) The new method applied to the problem in Fig.~\ref{fig:block-afqmc-infinite}. 
          The top row shows histograms of the expectation values in the 
          (a) denominator and (b) numerator of the 
          new ground-state energy estimator, compared with the CLT analysis.
          The middle row shows the respective variances,
          together with the computed error bars on 
          the variances, 
          versus sample size. 
          The purple dashed lines, which plot 
            $s/\sqrt{M}$, indicate the expected behavior of the error bars. 
            The bottom panel plots the size of the computed error bars on 
          the variances vs.~$1/\sqrt{M}$. The red solid lines show a linear fit, whose slopes give
          the values of $s$  above, for the denominator and numerator, respectively.
         } 
 \end{figure}

 In  Fig.~\ref{fig:block-afqmc-finite} we show results of the new method  applied to the example of 
 Fig.~\ref{fig:block-afqmc-infinite}. The histograms are shown for both the numerator and the denominator 
 in Eq.~(\ref{eq:soln-O-avg}) for the ground-state energy. 
 For brevity, results are only shown for one re-blocking size.
 It is seen that both approach perfect Gaussians in agreement with the CLT results. 
 The MC variances and the error bars on the variances are computed for both.
 The variances converge as  we increase the sample size, with the error bars on the
 variance decreasing as $1/\sqrt{M}$. 
 In other words, all the infinite variance symptoms have been removed.
 The behavior of the calculation is fundamentally different from before, and is consistent with 
 that of a finite, well-defined variance.

We next illustrate the problem and solution in calculations of 
 physical quantities besides the energy.
  A direct measure of magnetic order is 
 the spin-spin correlation function
 \begin{equation}
\textbf{S}_0 \cdot \textbf{S}_i = S_0^z S_i^z +\frac{1}{2} (S_0^+ S_i^- +S_0^- S_i^+),
\end{equation}
with 
$S_i^z=(n_{i\uparrow}-n_{i\downarrow})/2$,
$S_i^+ = c_{i\uparrow}^{\dagger}c_{i\downarrow}$, and $S_i^- = c_{i\downarrow}^{\dagger}c_{i\uparrow}$.
The site '0' is an arbitrary reference site and can be averaged over. The site $i$ is varied through the supercell, with its relative distance 
to site 0 denoted by $r$.
Thus far in the HS transformation
we have employed the spin decomposition, which is the more commonly adopted form in 
the repulsive Hubbard model.
Below we will  use the charge decomposition instead, which exhibits more severe symptoms of 
the infinite variance problem, to highlight the different features of the calculations 
{\it with\/} and {\it without\/} the  bridge link. The two sets of calculations will use
 otherwise identical settings, to compare  
 the computed spin-spin correlation functions.

\begin{figure}[htbp]
 \includegraphics[scale=0.40]{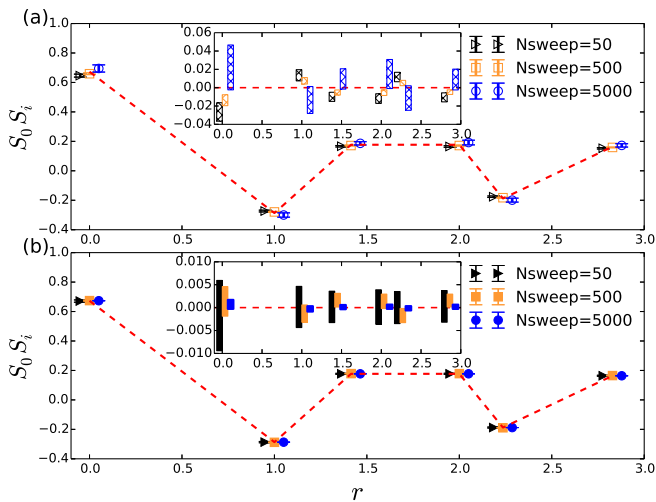}
 \caption{\label{fig:fix-example-4t4}
          (Color online) Comparison between the standard (top panel) and new (bottom panel) methods for spin correlations in the $4\times4$ Hubbard model with periodic boundary conditions and $U=8t$. 
          Results from exact diagonalization (ED) are shown by the red dashed lines. 
          The insets show the deviations from ED.
The three separate QMC results in each set, with increasing sample size,  
are displayed with small horizontal shifts for clarity. 
 In the new method the statistical error bars decrease as expected, while with the standard algorithm
 a drastic increase is seen 
in the largest run. Note that the vertical scale in inset (a) is five times that in inset (b).
         }
 \end{figure}
 
In Fig.~\ref{fig:fix-example-4t4}  we first show results
in a $4\times4$ system, where exact diagonalization (ED)
can be carried out for comparison. 
(The QMC calculations used a finite $\Delta\tau=0.01$ in units of $t$. 
The resulting Trotter error is negligible on the 
scale of the main plots. In the insets a shift has been applied to the ED 
results to account for it.) 
Within each panel, three QMC calculations are shown, with the number of independent 
measurements contained in the final result 
(denoted by the number of sweeps in each measurement block, Nsweep) 
 increasing by a factor of 10 every time. The CLT dictates that 
the statistical error should decrease by roughly $1/\sqrt{10}$ between the successive runs. 
In the standard algorithm (top panel), the computed error bar is seen to decrease first but to rise
dramatically in the largest run with Nsweep$=5,000$. 
(Note also the significantly higher than expected number of data points outside one error bar in the first 
two runs.)
The new method with bridge links eliminates the problem. The computed correlation functions are in agreement with exact results. The 
error bars decrease with increasing Nsweep as expected. 
In the run with $5,000$ sweeps, the results are about a factor of 30 more accurate than that from the 
standard algorithm. This would translate into, nominally, a factor of $\sim1,000$ savings in computing time. 
Of course the issue is much more fundamental than a quantitative gain, since 
the infinite variance means  
that the results from the standard algorithm 
cannot be assured of  
correctness within the context of its 
quoted error bars.

\begin{figure}[htbp]
 \includegraphics[scale=0.35]{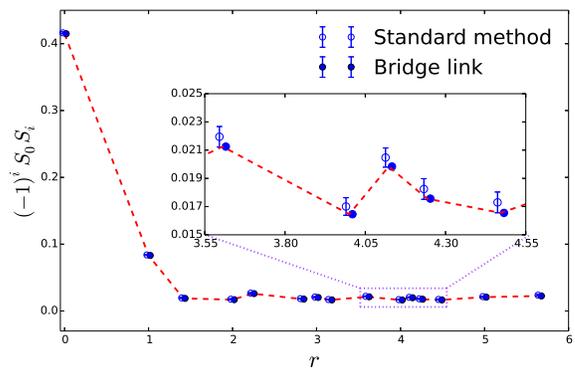}
 \caption{\label{fig:fix-example-8t8}
          (Color online) Comparison of the standard and new algorithms: spin correlations (staggered) in a larger lattice.
Results from the two sets of calculations are shown side-by-side, with a small horizontal shift for clarity. 
To aid the eye, those from the new method are connected by a red dashed line. 
          The system is an $8\times8$ Hubbard model   with periodic boundary condition and $U=0.5t$. 
         The inset shows a zoom of the segment indicated by the dotted purple
         rectangular box.
         }
 \end{figure}

In Fig.~\ref{fig:fix-example-8t8} we show results for a larger lattice. A smaller value of $U$ is studied, where
the antiferromagnetic order is weaker and higher accuracy is needed to resolve the order parameter (the magnetization, which can be thought of as the square root of the magnitude of the 
spin correlation at large distance). Once again, the results from the standard algorithm show large 
fluctuations.
The new approach removes the infinite variance problem, manifested for the chosen size runs as 
 a reduction in statistical error bar by a factor of 8.0.
The use of this new method has played an integral part in allowing us to obtain accurate results  at half-filling
for a variety of quantities,
and extrapolate reliably to the thermodynamic
limit \cite{2015arXiv150502290L}.

\section{Discussion}
\label{sec:discussion}

 The symptoms of the infinite variance problem tend to be subtle.
 We have observed that the calculation often  
 give ``reasonable'' results, i.e., the computed 
 expectation value is often
  in agreement with the correct answer 
 within (one or two of) the 
 estimated statistical error bar. 
 Different forms of HS transformations can show different levels of severity, as we further discuss below.
 Even within the same algorithm,  
different observables can behave differently. Further, 
the same observables can exhibit erratic behaviors in larger runs 
(more samples,
smaller time steps, longer imaginary-time lengths) 
which may have been masked in smaller ones.
 Perhaps the most common symptoms are occasional ``spikes'' among the 
 MC measurements of an observable, as illustrated earlier.  
 The behaviors seem consistent with a logarithm divergence of the variance. 
 If not controlled, the problem is likely to 
 manifest itself more strongly with growing computing power.
 More importantly, 
 the presence of an infinite variance means that,  in a strict mathematical sense, the results of all such simulations 
 are affected. 
 Without detailed 
 analysis or comparisons with properly controlled simulations, 
 one could not detect or predict which results may be biased or incorrect.

 Different HS transformations, which result in different forms of $\hat{B}$ in Eq.~(\ref{eq:Bx}),
 can lead to different behaviors of the determinantal QMC algorithm. 
 For example, in  the half-filled repulsive Hubbard model, both the
 charge (resulting in $\hat{B}(x) \propto \exp[i \gamma x  (n_{\i\uparrow}+n_{\i\downarrow})]$
 with $\gamma$ a real constant determined by $\Delta\tau$ and $U$)
 and spin  (resulting in $\hat{B}(x) \propto \exp[ \gamma x  (n_{\i\uparrow}-n_{\i\downarrow})]\,$)
 decompositions are free of the sign problem, as mentioned in Sec.~\ref{sec:formalism}. 
 Both lead to infinite variance problems but 
 the charge decomposition tends to have more severe symptoms. 
 The reason
 is that it yields an  imaginary form in the exponent,
 which causes the orbitals in
 the Slater determinants, upon propagation by $\hat{B}(x)$ (see Eq.~(\ref{eq:Thouless})), to acquire complex phases. 
 Although the overall integrand $f(\pmb{X})$ remains real and non-negative for any path $\pmb{X}$, the 
 random walks of the Slater determinants take place in the complex plane \cite{PhysRevLett.90.136401,AFQMC-lecture-notes-2013}, 
 rather than on the real axis 
 as with the spin decomposition. The ``two-dimensional'' nature of the random walks then causes the 
 density distribution of paths in the vicinity of  $f(\pmb{X})=0$ to tend to a finite value. This is 
 closely related to the general case where there is no symmetry protection 
 and a phase problem arises, for which a projection is necessary  \cite{PhysRevLett.90.136401,AFQMC-lecture-notes-2013}. 
 The finite density near $f(\pmb{X})=0$ exacerbates the divergence in Eq.~(\ref{eq:var}) and makes
 a more severe infinite variance problem.

 We comment that the infinite variance problem discussed here is absent in the constrained path Monte Carlo 
 \cite{PhysRevB.55.7464} or the phase-free AFQMC \cite{PhysRevLett.90.136401,AFQMC-lecture-notes-2013} methods, which are closely related to determinantal QMC.
 In the former approaches,  an importance sampling transformation is applied which modifies the propagator, 
 and thereby the PDF which is being sampled. 
 This is analogous to how the diffusion Monte Carlo (DMC) \cite{DMCCeperley,RevModPhys.73.33} approach works in fermion or other 
 systems in which the ground state wave function $\phi(R)$
 contains zeros (nodes). After importance sampling,
 one samples a distribution $\psi_T( R)\phi( R)$ which
 vanishes  quadratically where the trial wave function $\psi_T( R)=0$.
 (However some observables other than the energy can still have infinite variance \cite{moroni2014practical,Umrigar-variance}.) 
 The distinction between determinantal QMC and constrained path AFQMC is perhaps most easily 
 seen from the discussion and illustration in Fig.~1 of Ref.~\cite{PhysRevLett.83.2777}.
 When there is no sign problem, $P_l$ is non-negative, i.e., the region below the horizontal line of $P_l=0$ 
 is positive mirroring the region above, due to symmetry protection. 
 In determinantal QMC  all paths are sampled, while in constrained path only the 
 paths that stay exclusively above (or below) are sampled. In constrained path AFQMC, 
 the boundary condition and the importance sampling that imposes it cause the sampled 
 PDF to vanish  quadratically at   $P_l=0$, hence removing the infinite variance. 
 On the other hand, the answer from constrained path can be biased if 
 one uses a constraint which gives the incorrect  $P_l=0$.
 To remove the bias, one needs to modify the importance function so that the value 
 of  $\langle \psi_T|\phi\rangle$ is   ``lifted'' to be above zero, for example by adding a small constant
 similar to Eq.~(\ref{eq:fnew-shift}).
 The solution discussed in this paper,  using $\langle \psi_T|e^{-\Delta\tau \hat{H}}|\phi\rangle$, 
 provides a better way to do so. 
  
 To compute the intermediate function $\digamma(\pmb {X})$
in Eq.~(\ref{eq:soln-O-avg}), 
 we use the propagator
 written in the form $e^{-\Delta\tau \hat H}\doteq e^{-\Delta\tau \hat K/2}e^{-\Delta\tau \hat V}e^{-\Delta\tau \hat K/2}$.
 The two kinetic energy terms are first applied directly to $\langle \phi^{\rm l}|$ and $|\phi^{\rm r}\rangle$, respectively. With the resulting single determinants, 
 the interaction energy term, which  is expanded  in $\Delta\tau$
(to second order in most calculations), is computed in the usual way using the Green functions.
 For the $\hat{H}$'s studied in the present work, the interactions are local and the 
 second-order terms can be computed without significant increase in computational cost.
 Further improvements would be valuable for cases 
 with long-range interactions. 
 In principle, the $\Delta \tau$ in the propagator in $\digamma$ does not 
 have to have the same value as in the Trotter break-up in the rest of the simulation, although a 
 different value would make the ``bridge link'' static. 
 For example, one could use a smaller value of $\Delta \tau$ and place multiple ``bridge'' links 
 at fixed locations along the path where measurements will take place. 
 We have also tested the approach of evaluating the expectation value by directly applying Eq.~(\ref{eq:Bx}), sampling the 
 auxiliary-fields to evaluate the integral similar to the mixed estimator in constrained path AFQMC. 
 This can be used to complement  the power expansion approach when the overlap $f(\pmb {X})$ 
 is very small and a higher order expansion is needed.
 
 We have focused on ground-state calculations in our discussions. The ideas  apply
 to finite-temperature determinantal QMC as well. 
 In the standard grand-canonical algorithm \cite{assaad-lec-note,SANTOS2003}, the integrand 
 corresponding to  $f(\pmb {X})$ takes the form $\det[I+\prod_l B(x_l)]$, where $B$ is the matrix form 
 of the one-body propagator $\hat B$. 
 The structure of the path integrals and how $f(\pmb {X})$ varies with imaginary 
 time resemble closely \cite{PhysRevLett.83.2777} that of the ground-state projection,
 as we have invoked in the discussion above involving $P_l$. 
 When symmetry protection is present, $f(\pmb {X})$ becomes non-negative, however $f(\pmb {X})=0$ is in general 
not removed, since its removal would require the creation of 
a {\it finite\/} lower bound to  $f(\pmb {X})$ 
for any random choice of the path $\pmb {X}$ as the path length $l$ is increased. 
 We have carried out preliminary tests with the finite-temperature grand canonical algorithm \cite{BSS1PhysRevD.24.2278}, and 
 found  behaviors of the variance similar to those described in Sec.~\ref{sec:symptoms}.
 It is of course straightforward to apply the analysis we have discussed to 
 determine the presence of the infinite variance problem in any codes. 
 The simplest way to generalize the new algorithm to finite-temperature 
grand-canonical  determinantal QMC would be to set the  temperature and the chemical potential
 by $L$ time slices, and treat the bridge link only as a mathematical entity, although it will be worthwhile
 to study if other choices might be more efficient, especially near a phase transition.

 The infinite variance problem is not limited to sign-problem-free calculations. 
 In cases where the sign problem is present, one chooses to sample $|f(\pmb {X})|$ 
 and keep track of the sign in  
 evaluating Eq.~(\ref{eqn:integral}), so that the estimator in Eq.~(\ref{eq:Oavg}) is replaced by
 \begin{equation}
\langle \hat{O} \rangle\doteq \biggl \langle \frac{g(\pmb{X})}{f(\pmb{X})}\,s(\pmb{X})\biggr \rangle_{|f|}
\bigg/ \big \langle s(\pmb{X}) \big\rangle_{|f|}\,,
\label{eq:Oavg-sign}
\end{equation}
where 
 $s(\pmb {X})=f(\pmb {X})/|f(\pmb {X})|$.
 Because $f(\pmb {X})=0$ is not excluded in the PDF of $|f|$, the infinite variance problem will arise.
 In practice, the problem is
 entangled with the sign problem, which causes $\langle s\rangle$  to approach 
 zero --- and thus the statistically error to grow --- exponentially as $\beta$ or the system size is increased.
 As a result, the diverging variance can be obscured by 
 the
 large noise from the sign problem, especially for larger $\beta$ and system sizes. 
 However, 
 for a fixed  $\beta$
 and chosen system size, 
 the average sign $\langle s\rangle$  is {\it finite\/}. There is a well defined expectation value
 for the estimator above, and one would expect the MC error bar to converge  as $1/\sqrt{M}$ 
 with sample size. 
 The infinite variance problem causes a breakdown of this, in the same manner as in a sign-problem-free case. 
 One example where this point is relevant is in determinantal QMC as impurity solvers \cite{RevModPhys.68.13},
 where the finite-size of the cluster and the finite-temperatures help reduce the sign problem.

 There are additional areas where the 
 general ideas discussed in this paper can be useful.
 For example, in the presence of a sign problem, released node \cite{ceperley:releasednode} 
 calculations  in DMC or released constraint \cite{PhysRevB.88.125132,releasesorella} calculations in AFQMC 
 both require removing the zeros from the ``natural'' importance function 
 ($|\langle R|\psi_T\rangle|$ or $|\langle \psi_T|\phi\rangle|$). This is related to the issues described 
 here, and the bridge link approach, namely an importance function with an extra propagator inserted, can be an effective
 approach to generate the new importance function. More generally,
 the infinite variance problem can arise whenever the distribution being sampled, $f$, contains zeros where the 
 corresponding $g$ in denominator does not vanish. The analysis of the problem and the solution presented here will find use 
 in such situations which can occur in a variety of MC calculations, both quantum and classical.

\section{Conclusion}
\label{sec:discussion}

 Interacting quantum many-body systems form a central theme in many disciplines in physics,
 chemistry, and materials science. Because of their complexity and the high dimensionality of the Hilbert space involved, 
 Monte Carlo methods are often an indispensable tool in the study of such systems.  
 A Monte Carlo calculation computes an expectation value which inherently contains a statistical uncertainty. 
 Without a reliable estimate of the statistical error, the expectation value would become meaningless. 
 A divergence in the variance of the underlying many-dimensional integrals prevents the computation of 
 a reliable error bar, even in principle. 
 It is therefore vital
 to detect and then remedy this problem.
 This is the focal point of the present work.
  
 The determinantal QMC algorithms discussed in this paper are among the most widely applied
 computational approaches in physics. Determinantal QMC calculations are expected and assumed to provide  
 unbiased results in a variety 
 of otherwise intractable interacting fermion systems, which span multiple sub-disciplines of physics. 
 These results play a crucial role in our understanding of a variety of fundamental models and concepts.
 Recognizing that such calculations have an infinite variance problem and remedying it thus have 
 wide-ranging impacts. 
 The solution we have proposed removes the infinite variance problem in determinantal QMC,
 with simple 
 modifications 
 to the standard algorithms. The general ideas put forth are applicable in even broader contexts.

\section{Acknowledgments}

We thank Kostas Orginos, Mingpu Qin, Peter Rosenberg and Cyrus Umrigar for valuable discussions. This work was supported by 
NSF(grant no.~DMR-1409510) and the Simons Foundation. Computing was carried out at the computational
facilities at
William and Mary.

\bibliography{variance}
\end{document}